\documentclass{article}

\usepackage{amsmath}
\usepackage{amsfonts}
\usepackage{amsthm}
\usepackage{graphicx}
\usepackage{verbatim} 
\usepackage{bm}
\usepackage{bbm}
\usepackage{hyperref}
\usepackage{cite}
\usepackage{rotating}
\usepackage{epstopdf}
\usepackage{auto-pst-pdf}
\usepackage[margin=10pt,font={small,it},labelfont=bf,labelsep=endash]{caption}

\usepackage[top=1in, bottom=1.5in, left=1.5in, right=1.5in]{geometry}
\makeatletter
\renewcommand*{\@fnsymbol}[1]{\ifcase#1\or*\else\@arabic{\numexpr#1-1\relax}\fi}
\makeatother

\title{Approximate statistical alignment by iterative sampling of substitution matrices}
\author{
Joseph L.~Herman\thanks{Correspondence: {\tt herman@hms.harvard.edu}}~\,\thanks{Department of Statistics, University of Oxford; Center for Biomedical Informatics, Harvard Medical School.},
Adrienn Szab\'o\thanks{Institute of Computer Science and Control, Hungarian Academy of Sciences, Budapest.},
Istv\'an Mikl\'os\thanks{Institute of Computer Science and Control, Hungarian Academy of Sciences, Budapest; Department of Stochastics, Renyi Institute, Budapest.},
and Jotun Hein\thanks{Department of Statistics, University of Oxford.}\\
}

\begin{document}
\maketitle

\begin{abstract}
We outline a procedure for jointly sampling substitution matrices and multiple sequence alignments, according to an approximate posterior distribution, using an MCMC-based algorithm. This procedure provides an efficient and simple method by which to generate alternative alignments according to their expected accuracy, and allows appropriate parameters for substitution matrices to be selected in an automated fashion. In the cases considered here, the sampled alignments with the highest likelihood have an accuracy consistently higher than alignments generated using the standard BLOSUM62 matrix. 
\end{abstract}

\section*{Introduction}

Most commonly used sequence alignment programs (e.g. \cite{MAFFT,MUSCLE,CLUSTALW}) make use of a substitution matrix to specify the score associated with aligning different types of amino acids. Much work has been focused on the development of improved substitution matrices to improve the accuracy of the resulting alignments, and various algorithms have been developed to this end. 

One approach is to take a set of reference alignments, and to derive parameters that generate alignments that best match this reference set, either by matching the substitution parameters to observed statistics \cite{Dayhoff78,BLOSUM,Jones92,Muller02,Qiu06}, or by varying parameters in order to maximise the alignment accuracy with respect to the reference set \cite{Qian02,Hourai2004,Kim08,Huang2008,Edgar2009}. An alternative approach is to iteratively align the set of sequences, at each iteration deriving a new matrix from the observed pair frequencies within the aligned dataset \cite{Radivojac02}. 

While the early substitution matrices consisted of a small set of alternative matrices tailored for sequences of differing sequence identity \cite{Dayhoff78,BLOSUM}, later approaches allowed for these matrices to be adjusted to account for the amino acid content in a specific set of sequences \cite{Yu03}, as well as allowing for multiple classes of substitution matrix at different sites in the sequence \cite{Lartillot04}. Further improvements in accuracy have been obtained by accounting for of the primary \cite{Biegert09}, secondary \cite{Koshi95,Thorne96,Radivojac02,Szalkowski11} and tertiary \cite{Topham97,Rice97,Shi01} structural context of each residue, or by using matrices tailored for specific types of proteins \cite{Blundell93,Hill11}.

\subsection*{Impact of substitution matrix on alignment uncertainty}

An issue arising with all of the above approaches is that once the substitution matrix is selected, it is typically regarded as fixed for the purposes of subsequent analyses. Some algorithms include steps to select different substitution matrices at various stages of a progressive alignment \cite{CLUSTALW}, but these matrices are sampled from a fixed initial set. 

Since the resulting alignments will often be sensitive to the choice of matrix parameters, especially for more divergent sequences \cite{CLUSTALW}, the choice of substitution matrix may lead to significant bias in the resulting alignments. In some cases even small changes to the procedures used to derive optimal matrices may make a noticeable difference to the resulting alignment accuracy \cite{BLOSUM62_miscalculations}.

In addition, although procedures exist for adapting substitution matrices to sequences of interest \cite{Yu03,Yu05,Lartillot04}, the optimisation of the matrix parameters is generally carried out on the reference alignments rather than the sequences under analysis; unless explicitly accounted for, any differences in composition between datasets may be a further source of unpredictability.

Although several procedures have been devised for assessing the reliability of alignments (e.g. \cite{HoT,NOISY,GUIDANCE}), these procedures do not account for the uncertainty in the choice of substitution matrix, such that the reliability in the alignment will usually be overestimated. 

\subsection*{Joint sampling approaches}

One way to address this problem is to simultaneously sample substitution matrices and alignments from a joint posterior distribution, thereby incorporating parameter uncertainty into the alignment estimation. The approach of Zhu {\em et al.} \cite{Zhu98} focuses on pairwise alignments, and defines the space of possible substitution matrices as the BLOSUM or PAM series. In contrast, StatAlign is able to estimate parameters for arbitrary matrices once converted into rate matrix form, and does so while simultaneously sampling multiple sequence alignments and phylogenetic trees \cite{StatAlign}. However, this full joint sampling process is computationally intensive, limiting its application to larger numbers of sequences.

Here we explore an intermediate approach whereby substitution matrices are sampled from an approximate posterior distribution, and a single optimal alignment generated for each sampled matrix. This approach allows the effect of parameter uncertainty to be propagated into the alignment inference, while retaining much of the tractability of commonly used alignment algorithms.

\section*{Methods}

Substitution matrices are frequently defined in terms of log odds scores for pairwise homology statements. In the case of PAM matrices, these log odds scores are derived via conditional probabilities of one amino acid mutating into another within a particular time interval \cite{Dayhoff78}. BLOSUM matrices \cite{BLOSUM} are similarly based on the joint probabilities of observing a particular amino acid pairing in a set of reference alignments, with matrix entries defined as
\begin{equation}
  M_{a,b} = \frac{2}{\lambda} \log_2 \frac{p(a \diamond b)}{p(a)p(b)}
  \label{log_odds}
\end{equation}
where $a \diamond b$ denotes a homology statement between characters $a$ and $b$, and $p(a)$ denotes the background probability of character $a$. Rearranging and exponentiating, we obtain
\begin{equation}
p(a \diamond b) = p(a)p(b)\exp \left \{ \lambda M_{a,b} \right \}
\end{equation}
with $\lambda$ defined such that $\sum_{ab} p(a \diamond b) = 1$. As discussed by Yu {\em et al.} \cite{Yu03}, for a given substitution matrix, $M$, the matrix of pair probabilities can be uniquely recovered via the relationship ${P = Y^{-1}}$, where $Y_{ij} = 2^{\lambda M_{ij}}/2$.

While the pairing probabilities, $p(a\ \diamond\ b)$, are usually taken as fixed throughout the analysis, we proceed by conducting approximate posterior inference on these parameters.

\subsection*{Prior probability of substitution matrices}

We use the pair frequencies from the BLOCKS database \cite{BLOCKS} in order to construct an informative prior for the substitution matrix. To do so, we place the the following prior on the pair frequencies, centred around the observed frequencies used in the BLOSUM62 matrix
\begin{equation}
  f_{ij} \sim \textsf{Gamma}(\hat{f}_{ij}/\sigma^2,\hat{f}^2_{ij}/\sigma^2)
\label{freq_prior}
\end{equation}
which has mean $\hat{f}_{ij}$ and variance $\sigma^2$. The pairing probabilities are then derived as
\begin{equation}
p(i \diamond j) = \frac{f_{ij} }{ \sum_{ij} f_{ij}}
\end{equation}

\begin{sidewaysfigure}
  \includegraphics[width=1\linewidth]{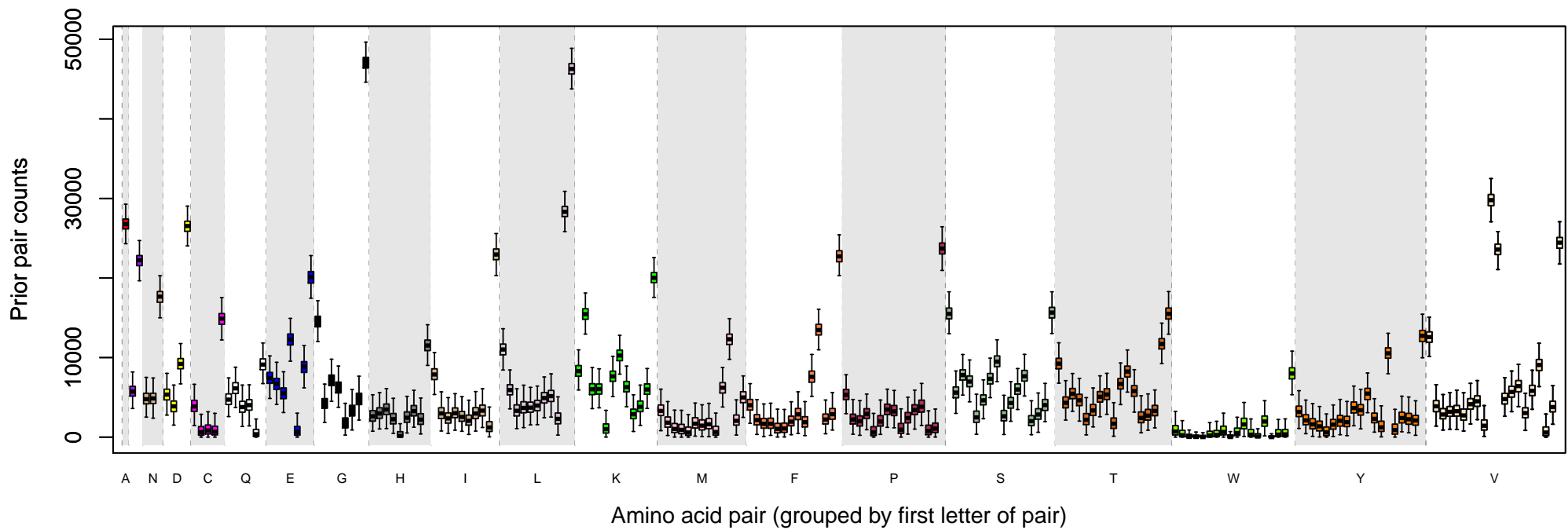}
\caption{\label{matrix_prior}The prior distribution on pair frequencies, shown for each amino acid pairing, as defined in equation~(\ref{freq_prior}). Each block of boxes (shown in the same colour, with boundaries highlighted by the alternating grey/white background) corresponds to pairings with the amino acid shown on the $x$-axis. Within each block, the ordering of the second letter of the pair follows the ordering of the first letters on the axis label. In each case the mean is centred around the observed frequencies taken from the counts in the BLOCKS database used to derive the BLOSUM62 matrix. Self pairings (corresponding to the right-most box in each group) typically show a much higher frequency, reflecting the level of sequence conservation in the database. Due to the symmetry of the matrix, only the upper triangle and diagonal elements are shown, such that moving from right to left across the plot involves progressively removing the first element of the group (for example, the first box in the 'V' block corresponds to a 'VA' amino acid pairing, whereas the first box in the 'Y' block corresponds to a 'YN' pairing, and the first box in the 'W' block to a 'WD' pairing).}
\end{sidewaysfigure}

One could in principle sample $\sigma^2$ from its posterior using MCMC, but in the current application we fix $\sigma=10^3$ (for comparison the standard deviation between the empirical frequencies is 6656.3), leading to a prior that moderately constrains the pair probabilities around the values in the original BLOSUM62 matrix, but allows for some significant variability, as shown in Figure~\ref{matrix_prior}.

\newgeometry{top=0.1in, bottom=0.1in, left=1.5in, right=1.5in}
\restoregeometry

\subsection*{Likelihood}

A sum-of-pairs alignment objective score is equivalent to the log likelihood under a Markov random field model, with independence between alignment columns. The overall likelihood of a set of sequences, $S$, given an alignment, $A$, and parameters, $\Theta$, can be written as 
\begin{align}
 p(S \mid A, \Theta) &\propto  \log \prod_{i=1}^L \prod_{j=1}^{K} p(A_{ji} \mid \Theta) \prod_{k=1}^{K-1} \prod_{l=k+1}^K \frac{p(A_{ki} \diamond A_{li}\mid \Theta)}{p(A_{ki}\mid \Theta)p(A_{li}\mid \Theta)}\\
\log p(S \mid A, \Theta)  &= \underbrace{\sum_{i=1}^L \sum_{j=1}^K \log p(A_{ji}\mid \Theta)}_{\mbox{sequence log likelihood}} + \underbrace{\sum_{i=1}^L \sum_{k=1}^{K-1} \sum_{l=k+1}^K \lambda M_{A_{ki},A_{li}}}_{\mbox{alignment score}}  + \mbox{const.}
\label{ali_score}
\end{align}
where $L$ is the number of columns in the alignment, $K$ is the number of sequences, and $A_{ki}$ represents either the character from sequence $k$ that is aligned to the $i$th column, or a gap character. The marginal probabilities $p(A_{ki}\mid \Theta)$ can also be modified to incorporate an affine gap penalty, such that
\begin{equation}
p(A_{ki} = -) = \left \{ \begin{array}{cc} 
g & \mbox{if } A_{k,i-1} = -\\
h & \mbox{if } A_{k,i-1} \neq -
\end{array} \right .
\end{equation} 
with $g < h$.

\subsection*{Approximate marginal likelihood}

As discussed earlier, full posterior sampling of multiple alignments is computationally intensive, such that joint estimation of substitution matrices and alignments is currently impractical on datasets with larger numbers of sequences.

In order to increase computational efficiency, we adopt an approximate procedure whereby the marginal likelihood of the sequences given a set of parameters $\Theta$ is assumed to be equal to the value of the likelihood corresponding to the alignment with the optimal score
\begin{equation}
 p(S \mid \Theta) \approx \sup_{A} p(S \mid A, \Theta)
\end{equation}
This approximation effectively asserts that the contribution of suboptimal alignments to the posterior is negligible, or, equivalently, that the prior on alignments for a particular $\Theta$ is a point mass at the maximum likelihood alignment. Under this approximation, alignment uncertainty is determined by the variance in $\Theta$.

\subsection*{Sampling according to expected alignment accuracy}

The search for alignments that maximise the target score is predicated on the assumption that the score is positively correlated with alignment accuracy for a given set of parameters. For pairwise alignments, the approximate distribution of alignment scores can be derived under various assumptions \cite{Karlin90}. However, for multiple sequences the objective scores used within the alignment programs may not carry sufficient information to predict alignment reliability \cite{Mitrophanov06}. As such, many alternative metrics have been developed for assessing alignment quality \cite{Thompson01,Lassmann05,Ahola08,DeBlasio12}. 

In this study, we consider a measure of alignment quality that is based on the number of non-gap homology statements in the alignment
\begin{equation}
q(A) = \sum_i (K-g_i(A))(K-g_i(A)-1)/2
\end{equation}
where $K$ is the number of sequences, and $g_i(A)$ denotes the number of gaps in column $i$ in alignment $A$.

As discussed in the Results section, empirically we observe a strong correlation between $q(A)$ and the alignment accuracy. For a random set of alignments, this quantity will not in general be correlated with the alignment accuracy, since it does not account for the sequence content. However, for optimal-scoring alignments generated according to equation~(\ref{ali_score}), any predicted homology statements must show significant evidence of being non-random in order to be included in the alignment, hence an alignment containing more homology statements should have a higher expected accuracy. 

\subsection*{Alternative likelihood function}

Given the empirically observed correlation between $q(A)$ and the alignment accuracy, we opt to sample substitution matrices according a modified marginal log likelihood of the form
\begin{equation}
\log \tilde{p}(S \mid \Theta) = \frac{1}{\tau(S)} q(\hat{A}[\Theta])
\label{alternative_likelihood}
\end{equation}
where $\hat{A}[\Theta] = \sup_A p(S \mid A, \Theta)$ is the optimal-scoring alignment under the original likelihood in equation~(\ref{ali_score}), and $\tau(S)$ is a measure of the variability in the alignment of the set of sequences $S$.

The quantity $q(A)$ can be roughly approximated as a sum of $N$ independent squared binomial variables with $n=K$ and $p=\chi/K$, where $\chi$ is the expected number of non-gap characters per column. Denoting the average sequence length by $\bar{L}$, then we must have $\chi N = K \bar{L}$, such that $N = \bar{L}/p$. Using the delta method, the approximate variance of each of these variables will be $\mathcal{O}(\chi^3(1-p))$. Hence, ignoring variability between the number of characters in each column, to a first approximation the variance of $q(A)$ for random alignments will be of the order of $(1-p)\chi^3N = \beta K^3 \bar{L}$, where $\beta = (1-p)p^2$, with a maximum value of $0.148$, when $p=2/3$. With a uniform prior on $\beta$, the posterior mean is approximately $0.08$. 

We therefore set $\tau$ to be the following function of the number of sequences, $K$, and the average sequence length, $\bar{L}$:
\begin{equation}
\tau(S) =  \sqrt{\beta K^3 \bar{L}} 
\label{tau}
\end{equation}
In principle one could sample $\beta$ from its posterior distribution, but the estimation of this quantity will be affected significantly by considering only a single alignment for each substitution matrix. In our analyses we fixed the value to $0.025$, below the approximate expected value for random alignments, but large enough to allow efficient traversal between modes of the likelihood. 

\subsection*{Alignment accuracy}

To measure the accuracy of generated alignments, we opt for the alignment metric accuracy (AMA) score introduced by Schwartz {\em et al.} \cite{Schwartz05}, since this possesses no inherent bias towards long or short alignments. To define the AMA score metric, we first define the following sets
\begin{align*}
  \begin{array}{rll}
  \mathcal{H}_H(A) &= \{(c_j,c_k) \mid \exists i \cdot (A_{ij} \neq -) , (A_{ik} \neq -)\}& \mbox{ pairwise homology statements}\\
  \mathcal{H}_D(A) &= \{(c_j,0) \mid \exists i \cdot (A_{ij} \neq -) , (A_{ik} = -)\}& \mbox{ pairwise deletions}\\
  \mathcal{H}_I(A) &= \{(0,c_k) \mid \exists i \cdot (A_{ij} = -) , (A_{ik} \neq -)\}& \mbox{ pairwise insertions}\\
  \mathcal{H}_N(A) &= \mathcal{H}_D(A) \cup \mathcal{H}_I(A) & \mbox{ pairwise non-homology statements}
  \end{array}
\end{align*}
where $c_j \in \{1,\hdots,|A_j|\}$, and $|A_j|$ is the length of the $j$th sequence. With these definitions, the accuracy of a predicted alignment, $P$, relative to the true alignment, $T$, is given by
\begin{align}
  a(P,T) &= a(T,P) = \frac{2 |\mathcal{H}_H(P) \cap \mathcal{H}_H(T)| + |\mathcal{H}_N(P) \cap \mathcal{H}_N(T)|}{(K-1) \sum_k |A_k|}
\end{align}

\subsection*{Varying the effective gap penalty}

Allowing the overall magnitude of the substitution matrix to vary is analogous to varying the gap parameters. Hence, by allowing the $\lambda$ parameter to vary, it is possible to simultaneously explore alternative substitution matrices and gap parameters, assuming a fixed ratio between the gap opening and extension penalties. 

\begin{figure}
  \centering
\includegraphics[width=\linewidth]{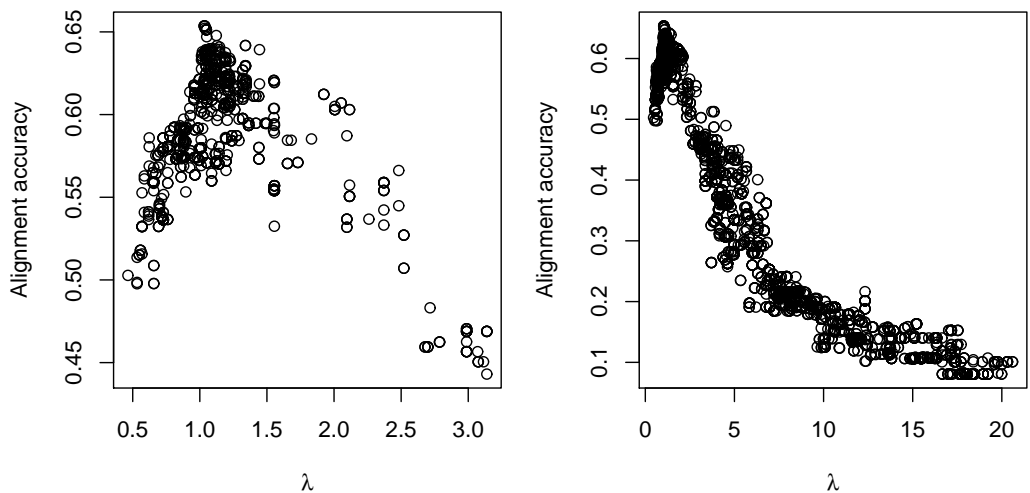}
\caption{\label{varying_lambda}When $\lambda$ is not constrained by an informative prior, it can become very large, since this increases the number of homology statements, thereby increasing the likelihood in equation~(\ref{alternative_likelihood}). However, beyond $\lambda=1.5$, this leads to a decrease in alignment accuracy, due to over-alignment of the sequences.}
\end{figure}

However, using a likelihood of the form in equation~(\ref{alternative_likelihood}), caution must be exercised when allowing $\lambda$ to vary, since large values of $\lambda$ will effectively render gaps exceedingly unlikely, causing the sequences to be over-aligned. As an example of this, Figure~\ref{varying_lambda} shows the posterior relationship between $\lambda$ and alignment accuracy when $\lambda$ is allowed to vary freely (i.e. with an uninformative prior), with gap parameters $g=-10$, $h=-1$. Up to a certain point (roughly $\lambda=1.5$), increasing $\lambda$ is associated with increased alignment accuracy, but beyond this the accuracy deteriorates rapidly, although the number of homology statements in the alignment continues to increase.

A similar pattern was observed in several other datasets (not shown), hence we opted to use an informative prior for $\lambda$ centred around unity, designed to keep $\lambda$ approximately within the range $[0.5,1.5]$.

\begin{equation}
p(\lambda) = \textsf{N}(\log \lambda \mid 0,0.1)
\end{equation}

\subsection*{MCMC scheme}

The general MCMC scheme we adopt consists of proposing a new set of pair frequencies, $f^\ast$, computing the corresponding log-odds substitution matrix, $M^\ast$, recomputing the optimal alignment for the new substitution matrix under the scoring scheme in equation~(\ref{ali_score}), and then accepting or rejecting the new matrix based on the equation~(\ref{alternative_likelihood}), i.e. accepting when

\begin{equation}
\log U(0,1) < \log \frac{\tilde{p}(S \mid M^\ast,g,h)}{\tilde{p}(S \mid M,g,h)} + \log\frac{p(f^\ast \mid \hat{f}, \sigma^2)}{p(f \mid \hat{f}, \sigma^2)}  
\label{accept_prob}
\end{equation}
where $\tilde{p}(\ \cdot \ )$ is as defined in equation~(\ref{alternative_likelihood}).

In order to improve mixing on the space of substitution matrices, we used two types of proposals. These involve taking subsets of matrix entries of size $n$, where $n=50,25$, and adding independent $U[-\rho_i,\rho_i]$ noise to the corresponding frequencies, where $i=1$ indicates $n=50$, and $i=2$ indicates $n=25$. By modifying submatrices of this size, on average between $90\%$ and $95\%$ of proposed changes to the substitution matrix lead to changes to the optimal alignment. 

In order to preserve the symmetry of the matrix, only the upper triangle and diagonal elements are modified, with the lower triangle updated accordingly afterwards. Since we are using symmetric proposal kernels, no adjustment to the Metropolis-Hastings ratio is required. Moves that lead to negative frequencies result in zero density under the prior, and hence are rejected.

The scaling factor, $\lambda$, is also sampled using a uniform random walk, characterised by a parameter $\rho_3$. 

The perturbation factors, $\rho_i$, are initialised at $70$, $10$ and $0.5$, and automatically tuned during the burn-in period, according to the following procedure: Every $10$ iterations, the acceptance rate for each move type is queried, and if it does not fall within the specified range (set by default to $[0.2,0.4]$ as per the considerations outlined by \cite{Roberts97}), the parameter $\rho_i$ is multiplied or divided by a factor of $0.9$, and the acceptance counts reset to zero. The different moves are selected according to weights $w_1=1, w_2=0.2$ and $w_3=0.5$.

\subsection*{Implementation details}

In general, finding the alignment that maximises a score of the type in equation~(\ref{ali_score}) is NP hard \cite{Wang94}. Multiple alignment programs generally make use of heuristic procedures such as progressive alignment in order to approximate this optimum.

In the current implementation of our iterative sampling scheme, we use the program MUSCLE \cite{MUSCLE} in order to generate the (approximately) optimal alignment for each substitution matrix.

Each sampled substitution matrix is written to file, and then an instance of MUSCLE is run using this substitution matrix, with the following command
\begin{verbatim}
muscle -gapopen g -gapextend h -matrix M -in seqs.fasta -out ali.fasta
\end{verbatim}
In order to increase sampling speed, we also used the flag {\tt -maxiters 2}, which restricts the number of refinement steps carried out by MUSCLE. Sampling $3000$ alignments after a burn-in of $1500$ iterations ($4500$ iterations overall) requires approximately $4.5$ minutes for a set of $15$ sequences, $6$ minutes for $33$ sequences, $9$ minutes for $60$ sequences, and $16$ minutes for $122$ sequences, increasing by a factor of approximately $1.5$ as the number of sequences is doubled. Significant improvements to the runtime could be obtained by incorporating the perturbation procedure directly into the MUSCLE code, such that intermediate read/write operations can be omitted from the workflow.

\begin{figure}
  \centering
\includegraphics[width=0.5\linewidth]{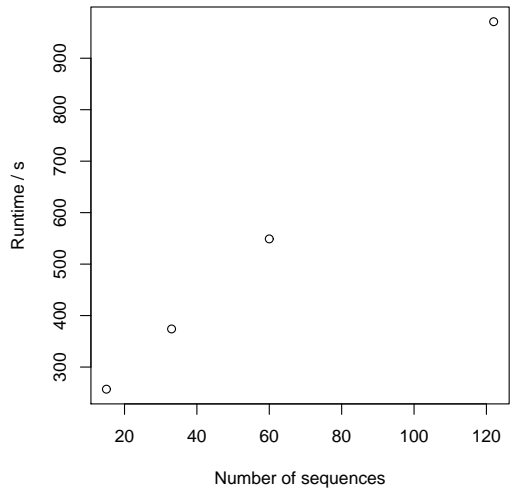}
\caption{\label{runtime}Time taken to generate $4500$ alignment samples under the iterative realignment MCMC algorithm, using MUSCLE to realign the sequences, on a single Intel i3 2.4GHz core. A significant portion of the runtime involves reading and writing to disk; the runtime could therefore be significantly reduced by integrating the perturbation procedure with the alignment code.}
\end{figure}

\section*{Results}

To evaluate the methodology, we conducted analyses on four datasets taken from the OXBench database \cite{Oxbench}. To generate these datasets, we selected one of the largest alignments from the OXBench suite [118], consisting of 122 immunoglobulin sequences, with average length 113. To assess how the alignment sampling method scaled with the number of sequences after controlling for other factors (such as amino acid content and sequence length), we subsampled smaller datasets from this alignment, yielding datasets with 15, 33, 60 and 122 sequences. These subsets were sampled so as to maximise dissimilarity within the subset, since the original alignment contained several well-defined subgroups that would otherwise skew the analysis. 

\begin{figure}
  \centering
\includegraphics[width=\linewidth]{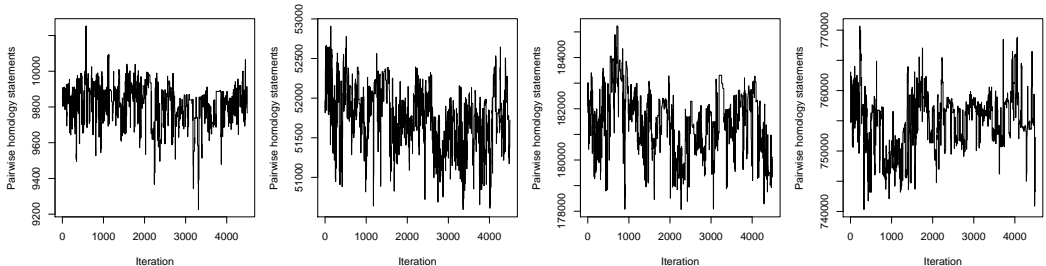}
\caption{\label{npairs_trace}Trace plot of the number of homology pairs in the alignment over the course of 4500 MCMC iterations, with a burn-in period of 1500 at the start, for $K=15, 33, 60, 122$ (left to right). This quantity plays the role of a log-likelihood in these simulations (after weighting by the factor $\tau$).}
\end{figure}

\begin{figure}
  \centering
\includegraphics[width=\linewidth]{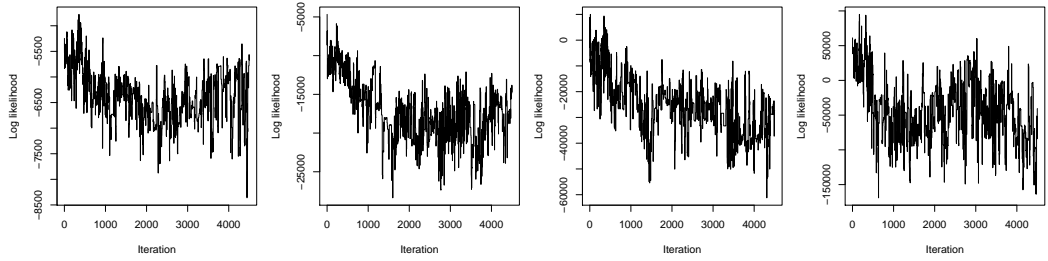}
\caption{\label{ll_trace}Trace plot of the original log likelihood in equation~(\ref{ali_score}) over the course of 4500 MCMC iterations, with a burn-in period of 1500 at the start, for $K=15, 33, 60, 122$ (left to right). In all cases, this quantity decreases substantially during the burn-in, since it is typically uncorrelated with the overall number of homology pairs in the alignment.}
\end{figure}

\begin{figure}
  \centering
\includegraphics[width=\linewidth]{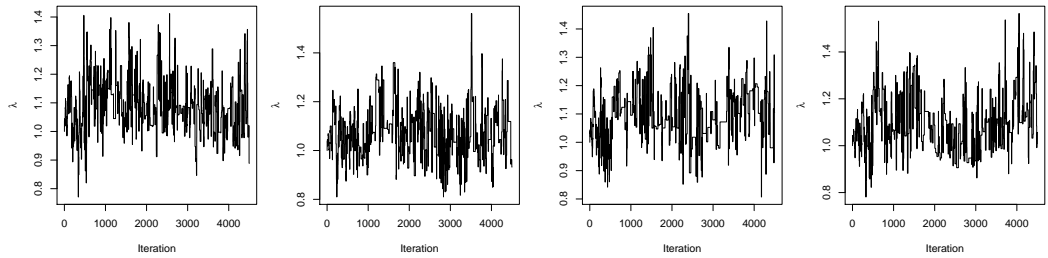}
\caption{\label{lambda_trace}Trace plot of the $\lambda$ parameter that acts as an inverse multiplier on all the entries of the substitution matrix, for $K=15, 33, 60, 122$ (left to right).}
\end{figure}

MCMC sampling was carried out for 4500 iterations, with the initial 1500 as burn-in. As shown by the trace plots in Figures~\ref{npairs_trace}-\ref{lambda_trace}, the key summary statistics appear to have successfully converged by the end of the burn-in. 

\begin{figure}
\includegraphics[width=\linewidth]{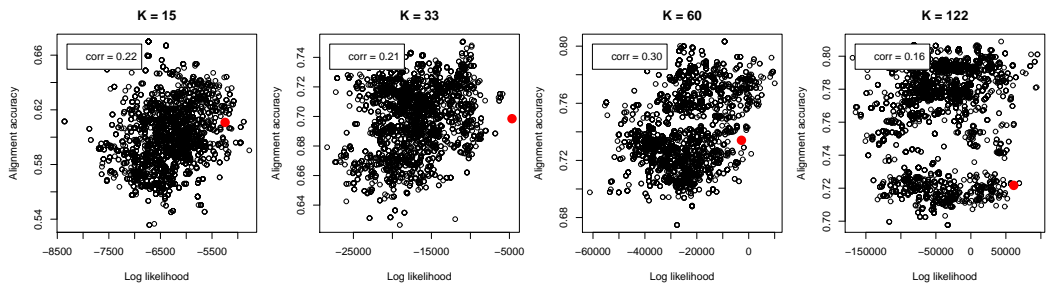}
\caption{\label{ll_accuracy}Although there is some correlation between log likelihood and alignment accuracy, it is generally weak. The alignment generated using the original BLOSUM62 matrix is shown in red.}
\end{figure}

The log likelihood shown in Figure~\ref{ll_trace} (corresponding to equation~(\ref{ali_score})) decreases during the burn-in. This quantity is not strongly correlated with the alignment accuracy ({\em see Figure~\ref{ll_accuracy}, Table~\ref{correlations}}), and is uncorrelated with the alternative likelihood used when accepting or rejecting the proposals according to equation~(\ref{accept_prob}) ($R^2 = 0.07, 0.18, -0.01, 0.01$ for $K=15,33,60,122$). Similarly, the multiplier $\lambda$ is also only weakly correlated with the alignment accuracy ({\em see Table~\ref{correlations}}).

\begin{figure}
\includegraphics[width=\linewidth]{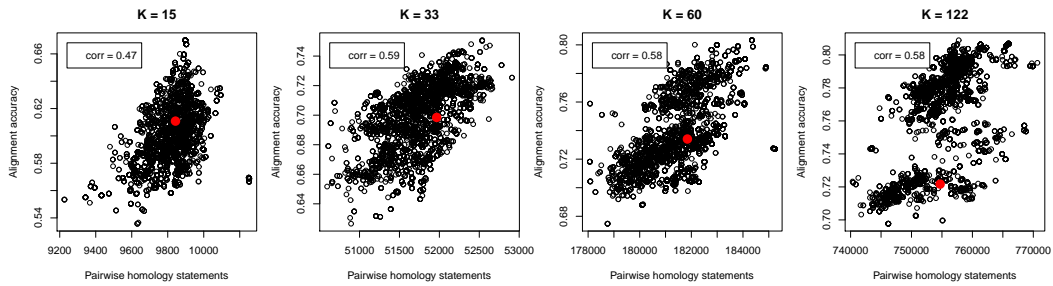}
\caption{\label{npairs_accuracy}The number of homology statements in the alignment is strongly correlated with the alignment accuracy. Although this will not be true in general for randomly generated alignments, for optimal alignments generated by a program such as MUSCLE, predicted homology statements have a higher probability of being accurate, hence larger numbers of homology statements is correlated with overall accuracy, justifying the form of the alternative likelihood in equation~(\ref{alternative_likelihood}). The alignment generated using the original BLOSUM62 matrix is shown in red.}
\end{figure}

In contrast, the number of homology pairs is strongly correlated with the alignment accuracy ({\em see Figure~\ref{npairs_accuracy}, Table~\ref{correlations}}), justifying the use of the alternative likelihood in equation~(\ref{alternative_likelihood}). 

\begin{figure}
\begin{center}
  \includegraphics[width=0.7\linewidth]{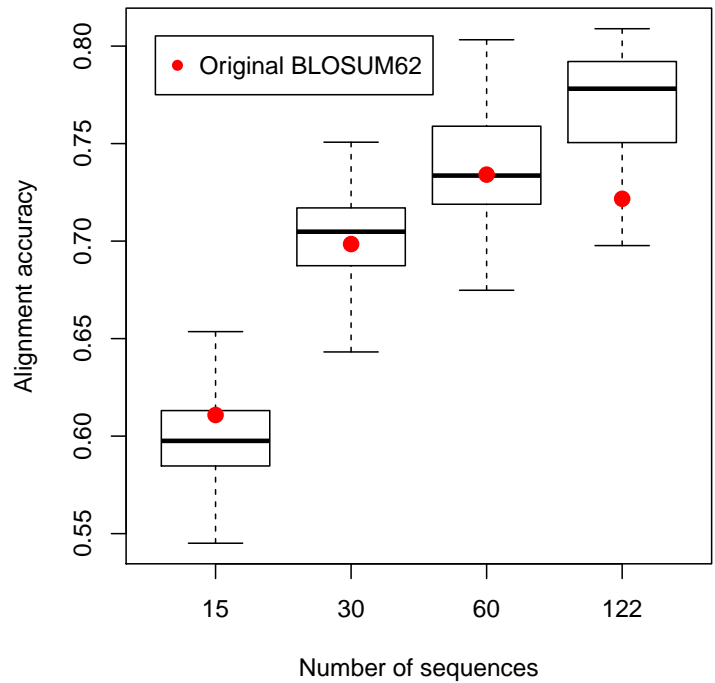}
\end{center}
\caption{\label{accuracy_boxplot}The distribution of alignment accuracy scores typically includes a large number of alignments with accuracy greater than that of the alignment derived from the initial BLOSUM62 matrix. As more sequences are added to the dataset, the average accuracy increases, due to additional information contained in the dataset. The variability remains roughly constant due to the dependency of $\tau$ on $K$ in equation~(\ref{tau}).}
\end{figure}

It is also notable that a significant proportion of the sampled alignments have accuracy higher than the alignment generated using the original BLOSUM62 matrix (shown in red in Figures~\ref{ll_accuracy}-\ref{npairs_accuracy} and Figure~\ref{accuracy_boxplot}).

On average, the sampled alignments are similar in length to the true alignments shorter (posterior mean lengths of 140, 149, 150 and 153, compared to the corresponding benchmark OXBench alignment lengths of 144, 150, 152 and 157 for $K=15,33,60,122$ respectively), while the alignments generated with the original BLOSUM62 matrix deviate more significantly in length when compared to the benchmarks (lengths 141, 141, 146 and 150 respectively).

\begin{center}
  \begin{table}
    \centering
  \begin{tabular}{cccc}
    \hline
    \# seqs & $R^2$ (log likelihood) & $R^2$ ($\lambda$) & $R^2$ (\# pairs)\\
    \hline
    15 & 0.22 & 0.23 & \textbf{0.47}\\
    33 & 0.21 & 0.36 & \textbf{0.59}\\
    60 & 0.30 & 0.20 & \textbf{0.58}\\
    122 & 0.16 & -0.01 & \textbf{0.58}\\
    \hline
  \end{tabular}
  \caption{\label{correlations}While the log likelihood of the optimal alignment has a weak positive correlation with the alignment accuracy, the number of pairwise homology statements is more strongly correlated with the accuracy (linear correlation coefficients shown in bold). The scaling factor $\lambda$ is also only weakly correlated with the accuracy. The log likelihood and number of pairs have a very low correlation with each other ($R^2 = 0.07, 0.18, -0.01, 0.01$ respectively).}
\end{table}
\end{center}

\subsection*{Small perturbations can have a large effect on the resulting alignment}

\begin{figure}
  \begin{center}
    \centering
    \includegraphics[width=0.7\linewidth]{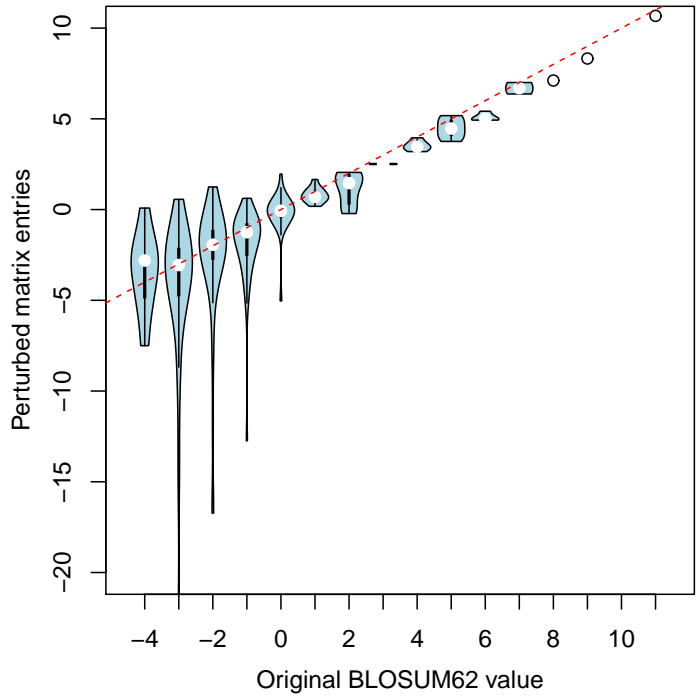}
  \end{center}
\caption{\label{matrix_comparison}A comparison of the matrix entries in the original BLOSUM62 matrix ($x$-axis) with the entries of the matrix yielding the highest accuracy alignment on the $60$-sequence dataset. While the distributions are centred around the initial values, there is significant variability, particularly for the negative entries (corresponding to pairings with a low probability).}
\end{figure}

Although the perturbations to the substitution matrix are generally small, constrained fairly strongly by the prior, these small modifications to the substitution matrix can often make a large difference to the resulting alignment. As an example, for the $60$-sequence OXBench dataset, while the original BLOSUM62 matrix in conjunction with MUSCLE yielded an alignment with AMA score of $0.73$, the $95\%$ highest posterior density interval spans the range $[0.70,0.79]$, and the maximum accuracy yielded by one of the sampled matrices is $0.80$. Figure~\ref{matrix_comparison} shows the distribution of entries of the matrix yielding this maximum accuracy alignment, illustrating how the entries are centred closely on the values in the original BLOSUM62 matrix, but with higher variance for the more negative entries (corresponding to low pairing probabilities). 

\subsection*{Posterior gap multiplier}

\begin{center}
\begin{figure}
  \centering
  \includegraphics[width=0.7\linewidth]{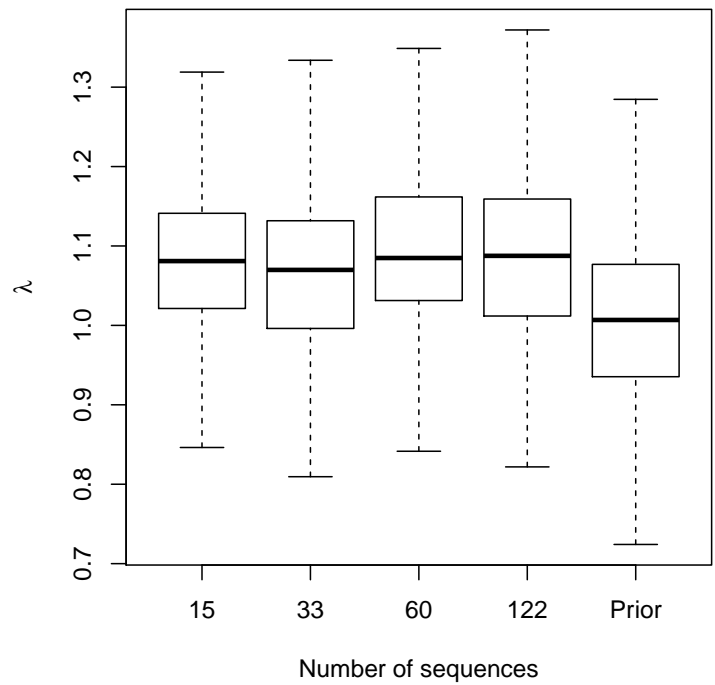}
\caption{\label{lambda_posterior}Posterior distributions for the $\lambda$ multiplier parameter, shown alongside the prior distribution (right).}
\end{figure}
\end{center}

As shown in Figure~\ref{lambda_trace}, the $\lambda$ multiplier parameter appears to converge and mix relatively successfully. Examining the posterior distribution of this quantity shows that it remains strongly constrained by the prior ({\em see Figure~\ref{lambda_posterior}}), but shifted slightly upwards towards higher values of $\lambda$.

\section*{Conclusions}

The method outlined here provides a simple approach for incorporating parameter uncertainty into score-based alignment, generating a set of alternative multiple alignments for a set of sequences, and sampling substitution matrices based upon an approximate likelihood that is a good predictor of the accuracy of the resulting alignments. 

In the examples considered here, varying the substitution matrix parameters can significantly affect alignment accuracy, and the high-likelihood alignments typically have a higher accuracy than the alignment generated using the standard BLOSUM substitution matrix.

As well as providing a way to address parameter uncertainty, this approach provides a way of systematically generating alignments according to a probability distribution that is correlated with alignment accuracy. These sampled alignments may provide a useful starting point for assessing the affect of alignment uncertainty downstream analyses \cite{WeaveAlign}.

As discussed by Herman {\em et al.}, the minimum-risk summary alignment derived from a set of posterior alignment samples typically is more accurate than the majority of the individual samples \cite{WeaveAlign}. The minimum-risk summary procedure could be used in conjunction with the sampling procedure outlined here in order to generate more reliable alignments.

More extensive tests of this procedure on alignment benchmark databases are required to determine whether this observed relationship between number of homology pairs and alignment accuracy holds in a more general context.

\section*{Additional material}
BLOSUM matrices and pair frequencies were downloaded from
\begin{verbatim}
ftp://ftp.ncbi.nih.gov/repository/blocks/unix/blosum/blosum.tar.Z
\end{verbatim}
(last accesed 18 January, 2015).

\bibliography{iterative_sampling}


\begin{thebibliography}{10}
\providecommand{\url}[1]{[#1]}
\providecommand{\urlprefix}{}

\bibitem{MAFFT}
Katoh K, Kuma K, Toh H, Miyata T: \textbf{MAFFT version 5: improvement in
  accuracy of multiple sequence alignment}. \emph{Nucleic Acids Res} 2005,
  \textbf{33}:511--8.

\bibitem{MUSCLE}
Edgar R: \textbf{MUSCLE: multiple sequence alignment with high accuracy and
  high throughput}. \emph{Nucleic Acids Res} 2004, \textbf{32}:1792--7.

\bibitem{CLUSTALW}
Thompson JD, Higgins DG, Gibson TJ: \textbf{CLUSTAL W: improving the
  sensitivity of progressive multiple sequence alignment through sequence
  weighting, position-specific gap penalties and weight matrix choice}.
  \emph{Nucleic acids research} 1994, \textbf{22}(22):4673--4680.

\bibitem{Dayhoff78}
Dayhoff MO, Schwartz RM, Orcutt BC: \textbf{A model of evolutionary change in
  proteins}. \emph{Atlas of Protein Sequence and Structure} 1978,
  \textbf{5}(suppl 3):345--351.

\bibitem{BLOSUM}
Henikoff S, Henikoff JG: \textbf{Amino acid substitution matrices from protein
  blocks}. \emph{Proceedings of the National Academy of Sciences} 1992,
  \textbf{89}(22):10915--10919.

\bibitem{Jones92}
Jones DT, Taylor WR, Thornton JM: \textbf{The rapid generation of mutation data
  matrices from protein sequences}. \emph{Computer applications in the
  biosciences: CABIOS} 1992, \textbf{8}(3):275--282.

\bibitem{Muller02}
Muller T, Spang R, Vingron M: \textbf{Estimating amino acid substitution
  models: a comparison of Dayhoff's estimator, the resolvent approach and a
  maximum likelihood method}. \emph{Mol Biol Evol} 2002, \textbf{19}:8--13.

\bibitem{Qiu06}
Qiu J, Elber R: \textbf{SSALN: An alignment algorithm using structure-dependent
  substitution matrices and gap penalties learned from structurally aligned
  protein pairs}. \emph{Proteins: Structure, Function, and Bioinformatics}
  2006, \textbf{62}(4):881--891.

\bibitem{Qian02}
Qian B, Goldstein RA: \textbf{Optimization of a new score function for the
  generation of accurate alignments}. \emph{Proteins: Structure, Function, and
  Bioinformatics} 2002, \textbf{48}(4):605--610.

\bibitem{Hourai2004}
Hourai Y, Akutsu T, Akiyama Y: \textbf{Optimizing substitution matrices by
  separating score distributions}. \emph{Bioinformatics} 2004,
  \textbf{20}(6):863--873.

\bibitem{Kim08}
Kim E, Kececioglu J: \textbf{Learning scoring schemes for sequence alignment
  from partial examples}. \emph{IEEE/ACM Transactions on Computational Biology
  and Bioinformatics (TCBB)} 2008, \textbf{5}(4):546--556.

\bibitem{Huang2008}
Huang X: \textbf{Sequence alignment with an appropriate substitution matrix}.
  \emph{Journal of Computational Biology} 2008, \textbf{15}(2):129--138.

\bibitem{Edgar2009}
Edgar R: \textbf{Optimizing substitution matrix choice and gap parameters for
  sequence alignment}. \emph{BMC Bioinformatics} 2009, \textbf{10}:396.

\bibitem{Radivojac02}
Radivojac P, Obradovic Z, Brown CJ, Dunker AK: \textbf{Improving sequence
  alignments for intrinsically disordered proteins.} In \emph{Pacific Symposium
  on Biocomputing}, \emph{Volume~7} 2002:589--600.

\bibitem{Yu03}
Yu YK, Wootton JC, Altschul SF: \textbf{The compositional adjustment of amino
  acid substitution matrices}. \emph{Proceedings of the National Academy of
  Sciences} 2003, \textbf{100}(26):15688--15693.

\bibitem{Lartillot04}
Lartillot N, Philippe H: \textbf{A Bayesian Mixture Model for Across-Site
  Heterogeneities in the Amino-Acid Replacement Process}. \emph{Molecular
  Biology and Evolution} 2004, \textbf{21}(6):1095--1109.

\bibitem{Biegert09}
Biegert A, S{\"o}ding J: \textbf{Sequence context-specific profiles for
  homology searching}. \emph{Proceedings of the National Academy of Sciences}
  2009, \textbf{106}(10):3770--3775.

\bibitem{Koshi95}
Koshi JM, Goldstein RA: \textbf{Context-dependent optimal substitution
  matrices}. \emph{Protein Engineering} 1995, \textbf{8}(7):641--645.

\bibitem{Thorne96}
Thorne JL, Goldman N, Jones DT: \textbf{Combining protein evolution and
  secondary structure.} \emph{Molecular Biology and Evolution} 1996,
  \textbf{13}(5):666--673.

\bibitem{Szalkowski11}
Szalkowski AM, Anisimova M: \textbf{Markov Models of Amino Acid Substitution to
  Study Proteins with Intrinsically Disordered Regions}. \emph{PLoS ONE} 2011,
  \textbf{6}(5):e20488.

\bibitem{Topham97}
Topham CM, Srinivasan N, Blundell TL: \textbf{Prediction of the stability of
  protein mutants based on structural environment-dependent amino acid
  substitution and propensity tables.} \emph{Protein Engineering} 1997,
  \textbf{10}:7--21.

\bibitem{Rice97}
Rice DW, Eisenberg D: \textbf{A 3D-1D substitution matrix for protein fold
  recognition that includes predicted secondary structure of the sequence}.
  \emph{Journal of Molecular Biology} 1997, \textbf{267}(4):1026 -- 1038.

\bibitem{Shi01}
Shi J, Blundell TL, Mizuguchi K: \textbf{FUGUE: sequence-structure homology
  recognition using environment-specific substitution tables and
  structure-dependent gap penalties}. \emph{Journal of Molecular Biology} 2001,
  \textbf{310}:243 -- 257.

\bibitem{Blundell93}
Blundell TL, Donnelly D, Overington JP, Ruffle SV, Nugent JH: \textbf{Modeling
  $\alpha$-helical transmembrane domains: The calculation and use of
  substitution tables for lipid-facing residues}. \emph{Protein Science} 1993,
  \textbf{2}:55--70.

\bibitem{Hill11}
Hill JR, Kelm S, Shi J, Deane CM: \textbf{Environment specific substitution
  tables improve membrane protein alignment}. \emph{Bioinformatics} 2011,
  \textbf{27}(13):i15--i23.

\bibitem{BLOSUM62_miscalculations}
Styczynski MP, Jensen KL, Rigoutsos I, Stephanopoulos G: \textbf{BLOSUM62
  miscalculations improve search performance}. \emph{Nature biotechnology}
  2008, \textbf{26}(3):274--275.

\bibitem{Yu05}
Yu YK, Altschul SF: \textbf{The construction of amino acid substitution
  matrices for the comparison of proteins with non-standard compositions}.
  \emph{Bioinformatics} 2005, \textbf{21}(7):902--911.

\bibitem{HoT}
Landan G, Graur D: \textbf{Heads or Tails: A simple reliability check for
  multiple sequence alignments}. \emph{Molecular Biology and Evolution} 2007,
  \textbf{24}(6):1380--1383.

\bibitem{NOISY}
Dress A, Flamm C, Fritzsch G, Grunewald S, Kruspe M, Prohaska S, Stadler P:
  \textbf{Noisy: Identification of problematic columns in multiple sequence
  alignments}. \emph{Algorithms for Molecular Biology} 2008, \textbf{3}:7.

\bibitem{GUIDANCE}
Penn O, Privman E, Landan G, Graur D, Pupko T: \textbf{An alignment confidence
  score capturing robustness to guide tree uncertainty}. \emph{Molecular
  Biology and Evolution} 2010, \textbf{27}(8):1759--1767.

\bibitem{Zhu98}
Zhu J, Liu JS, Lawrence CE: \textbf{Bayesian adaptive sequence alignment
  algorithms.} \emph{Bioinformatics} 1998, \textbf{14}:25--39.

\bibitem{StatAlign}
Nov\'ak A, Mikl\'os I, Lyngs\o{} R, Hein J: \textbf{{StatAlign: an extendable
  software package for joint Bayesian estimation of alignments and evolutionary
  trees}}. \emph{Bioinformatics} 2008, \textbf{24}(20):2403--2404.

\bibitem{BLOCKS}
Henikoff S, Henikoff JG, Pietrokovski S: \textbf{Blocks+: a non-redundant
  database of protein alignment blocks derived from multiple compilations.}
  \emph{Bioinformatics} 1999, \textbf{15}(6):471--479.

\bibitem{Karlin90}
Karlin S, Altschul SF: \textbf{Methods for assessing the statistical
  significance of molecular sequence features by using general scoring
  schemes}. \emph{Proceedings of the National Academy of Sciences} 1990,
  \textbf{87}(6):2264--2268.

\bibitem{Mitrophanov06}
Mitrophanov AY, Borodovsky M: \textbf{Statistical significance in biological
  sequence analysis}. \emph{Briefings in Bioinformatics} 2006,
  \textbf{7}:2--24.

\bibitem{Thompson01}
Thompson JD, Plewniak F, Ripp R, Thierry JC, Poch O: \textbf{Towards a reliable
  objective function for multiple sequence alignments}. \emph{Journal of
  Molecular Biology} 2001, \textbf{314}(4):937 -- 951.

\bibitem{Lassmann05}
Lassmann T, Sonnhammer ELL: \textbf{Automatic assessment of alignment quality}.
  \emph{Nucleic Acids Research} 2005, \textbf{33}(22):7120--7128.

\bibitem{Ahola08}
Ahola V, Aittokallio T, Vihinen M, Uusipaikka E: \textbf{Model-based prediction
  of sequence alignment quality}. \emph{Bioinformatics} 2008,
  \textbf{24}(19):2165--2171.

\bibitem{DeBlasio12}
DeBlasio D, Wheeler T, Kececioglu J: \textbf{Estimating the accuracy of
  multiple alignments and its use in parameter advising}. In \emph{Research in
  Computational Molecular Biology}, \emph{Volume 7262 of \emph{Lecture Notes in
  Computer Science}}. Edited by Chor B, Springer Berlin Heidelberg 2012:45--59.

\bibitem{Schwartz05}
Schwartz AS, Myers EW, Pachter L: \textbf{Alignment metric accuracy}.
  \emph{arXiv:q-bio/0510052} 2005.

\bibitem{Roberts97}
Roberts GO, Gelman A, Gilks WR: \textbf{Weak convergence and optimal scaling of
  random walk {Metropolis} algorithms}. \emph{Annals of Applied Probability}
  1997, \textbf{7}:110--120.

\bibitem{Wang94}
Wang L, Jiang T: \textbf{On the complexity of multiple sequence alignment}.
  \emph{Journal of Computational Biology} 1994, \textbf{1}(4):337--348.

\bibitem{Oxbench}
Raghava G, Searle S, Audley P, Barber J, Barton G: \textbf{OXBench: A benchmark
  for evaluation of protein multiple sequence alignment accuracy}. \emph{BMC
  Bioinformatics} 2003, \textbf{4}:47.

\bibitem{WeaveAlign}
Herman JL, Nov\'ak A, Lyngs\o{} R, Szab\'o A, Mikl\'os I, Hein J:
  \textbf{Efficient representation of uncertainty in multiple sequence
  alignments using directed acyclic graphs}. \emph{(submitted)} 2014.

\end{thebibliography}

\newcommand{\BMCxmlcomment}[1]{}

\BMCxmlcomment{

<refgrp>

<bibl id="B1">
  <title><p>MAFFT version 5: improvement in accuracy of multiple sequence
  alignment</p></title>
  <aug>
    <au><snm>Katoh</snm><fnm>K</fnm></au>
    <au><snm>Kuma</snm><fnm>K</fnm></au>
    <au><snm>Toh</snm><fnm>H</fnm></au>
    <au><snm>Miyata</snm><fnm>T</fnm></au>
  </aug>
  <source>Nucleic Acids Res</source>
  <pubdate>2005</pubdate>
  <volume>33</volume>
  <fpage>511</fpage>
  <lpage>8</lpage>
</bibl>

<bibl id="B2">
  <title><p>MUSCLE: multiple sequence alignment with high accuracy and high
  throughput</p></title>
  <aug>
    <au><snm>Edgar</snm><fnm>RC</fnm></au>
  </aug>
  <source>Nucleic Acids Res</source>
  <pubdate>2004</pubdate>
  <volume>32</volume>
  <fpage>1792</fpage>
  <lpage>7</lpage>
</bibl>

<bibl id="B3">
  <title><p>CLUSTAL W: improving the sensitivity of progressive multiple
  sequence alignment through sequence weighting, position-specific gap
  penalties and weight matrix choice</p></title>
  <aug>
    <au><snm>Thompson</snm><fnm>JD</fnm></au>
    <au><snm>Higgins</snm><fnm>DG</fnm></au>
    <au><snm>Gibson</snm><fnm>TJ</fnm></au>
  </aug>
  <source>Nucleic acids research</source>
  <publisher>Oxford Univ Press</publisher>
  <pubdate>1994</pubdate>
  <volume>22</volume>
  <issue>22</issue>
  <fpage>4673</fpage>
  <lpage>-4680</lpage>
</bibl>

<bibl id="B4">
  <title><p>A model of evolutionary change in proteins</p></title>
  <aug>
    <au><snm>Dayhoff</snm><fnm>M. O.</fnm></au>
    <au><snm>Schwartz</snm><fnm>R. M.</fnm></au>
    <au><snm>Orcutt</snm><fnm>B. C.</fnm></au>
  </aug>
  <source>Atlas of Protein Sequence and Structure</source>
  <pubdate>1978</pubdate>
  <volume>5</volume>
  <issue>suppl 3</issue>
  <fpage>345</fpage>
  <lpage>351</lpage>
</bibl>

<bibl id="B5">
  <title><p>Amino acid substitution matrices from protein blocks</p></title>
  <aug>
    <au><snm>Henikoff</snm><fnm>S</fnm></au>
    <au><snm>Henikoff</snm><fnm>JG</fnm></au>
  </aug>
  <source>Proceedings of the National Academy of Sciences</source>
  <publisher>National Acad Sciences</publisher>
  <pubdate>1992</pubdate>
  <volume>89</volume>
  <issue>22</issue>
  <fpage>10915</fpage>
  <lpage>-10919</lpage>
</bibl>

<bibl id="B6">
  <title><p>The rapid generation of mutation data matrices from protein
  sequences</p></title>
  <aug>
    <au><snm>Jones</snm><fnm>DT</fnm></au>
    <au><snm>Taylor</snm><fnm>WR</fnm></au>
    <au><snm>Thornton</snm><fnm>JM</fnm></au>
  </aug>
  <source>Computer applications in the biosciences: CABIOS</source>
  <publisher>Oxford Univ Press</publisher>
  <pubdate>1992</pubdate>
  <volume>8</volume>
  <issue>3</issue>
  <fpage>275</fpage>
  <lpage>-282</lpage>
</bibl>

<bibl id="B7">
  <title><p>Estimating amino acid substitution models: a comparison of
  Dayhoff's estimator, the resolvent approach and a maximum likelihood
  method</p></title>
  <aug>
    <au><snm>Muller</snm><fnm>T</fnm></au>
    <au><snm>Spang</snm><fnm>R</fnm></au>
    <au><snm>Vingron</snm><fnm>M</fnm></au>
  </aug>
  <source>Mol Biol Evol</source>
  <pubdate>2002</pubdate>
  <volume>19</volume>
  <fpage>8</fpage>
  <lpage>13</lpage>
</bibl>

<bibl id="B8">
  <title><p>SSALN: An alignment algorithm using structure-dependent
  substitution matrices and gap penalties learned from structurally aligned
  protein pairs</p></title>
  <aug>
    <au><snm>Qiu</snm><fnm>J</fnm></au>
    <au><snm>Elber</snm><fnm>R</fnm></au>
  </aug>
  <source>Proteins: Structure, Function, and Bioinformatics</source>
  <publisher>Wiley Online Library</publisher>
  <pubdate>2006</pubdate>
  <volume>62</volume>
  <issue>4</issue>
  <fpage>881</fpage>
  <lpage>-891</lpage>
</bibl>

<bibl id="B9">
  <title><p>Optimization of a new score function for the generation of accurate
  alignments</p></title>
  <aug>
    <au><snm>Qian</snm><fnm>B</fnm></au>
    <au><snm>Goldstein</snm><fnm>RA</fnm></au>
  </aug>
  <source>Proteins: Structure, Function, and Bioinformatics</source>
  <publisher>Wiley Online Library</publisher>
  <pubdate>2002</pubdate>
  <volume>48</volume>
  <issue>4</issue>
  <fpage>605</fpage>
  <lpage>-610</lpage>
</bibl>

<bibl id="B10">
  <title><p>Optimizing substitution matrices by separating score
  distributions</p></title>
  <aug>
    <au><snm>Hourai</snm><fnm>Y</fnm></au>
    <au><snm>Akutsu</snm><fnm>T</fnm></au>
    <au><snm>Akiyama</snm><fnm>Y</fnm></au>
  </aug>
  <source>Bioinformatics</source>
  <publisher>Oxford Univ Press</publisher>
  <pubdate>2004</pubdate>
  <volume>20</volume>
  <issue>6</issue>
  <fpage>863</fpage>
  <lpage>-873</lpage>
</bibl>

<bibl id="B11">
  <title><p>Learning scoring schemes for sequence alignment from partial
  examples</p></title>
  <aug>
    <au><snm>Kim</snm><fnm>E</fnm></au>
    <au><snm>Kececioglu</snm><fnm>J</fnm></au>
  </aug>
  <source>IEEE/ACM Transactions on Computational Biology and Bioinformatics
  (TCBB)</source>
  <publisher>IEEE Computer Society Press</publisher>
  <pubdate>2008</pubdate>
  <volume>5</volume>
  <issue>4</issue>
  <fpage>546</fpage>
  <lpage>-556</lpage>
</bibl>

<bibl id="B12">
  <title><p>Sequence alignment with an appropriate substitution
  matrix</p></title>
  <aug>
    <au><snm>Huang</snm><fnm>X</fnm></au>
  </aug>
  <source>Journal of Computational Biology</source>
  <publisher>Mary Ann Liebert, Inc. 2 Madison Avenue Larchmont, NY 10538
  USA</publisher>
  <pubdate>2008</pubdate>
  <volume>15</volume>
  <issue>2</issue>
  <fpage>129</fpage>
  <lpage>-138</lpage>
</bibl>

<bibl id="B13">
  <title><p>Optimizing substitution matrix choice and gap parameters for
  sequence alignment</p></title>
  <aug>
    <au><snm>Edgar</snm><fnm>R</fnm></au>
  </aug>
  <source>BMC Bioinformatics</source>
  <pubdate>2009</pubdate>
  <volume>10</volume>
  <issue>1</issue>
  <fpage>396</fpage>
</bibl>

<bibl id="B14">
  <title><p>Improving sequence alignments for intrinsically disordered
  proteins.</p></title>
  <aug>
    <au><snm>Radivojac</snm><fnm>P</fnm></au>
    <au><snm>Obradovic</snm><fnm>Z</fnm></au>
    <au><snm>Brown</snm><fnm>CJ</fnm></au>
    <au><snm>Dunker</snm><fnm>AK</fnm></au>
  </aug>
  <source>Pacific Symposium on Biocomputing</source>
  <pubdate>2002</pubdate>
  <volume>7</volume>
  <fpage>589</fpage>
  <lpage>-600</lpage>
</bibl>

<bibl id="B15">
  <title><p>The compositional adjustment of amino acid substitution
  matrices</p></title>
  <aug>
    <au><snm>Yu</snm><fnm>YK</fnm></au>
    <au><snm>Wootton</snm><fnm>JC</fnm></au>
    <au><snm>Altschul</snm><fnm>SF</fnm></au>
  </aug>
  <source>Proceedings of the National Academy of Sciences</source>
  <publisher>National Acad Sciences</publisher>
  <pubdate>2003</pubdate>
  <volume>100</volume>
  <issue>26</issue>
  <fpage>15688</fpage>
  <lpage>-15693</lpage>
</bibl>

<bibl id="B16">
  <title><p>A Bayesian Mixture Model for Across-Site Heterogeneities in the
  Amino-Acid Replacement Process</p></title>
  <aug>
    <au><snm>Lartillot</snm><fnm>N</fnm></au>
    <au><snm>Philippe</snm><fnm>H</fnm></au>
  </aug>
  <source>Molecular Biology and Evolution</source>
  <pubdate>2004</pubdate>
  <volume>21</volume>
  <issue>6</issue>
  <fpage>1095</fpage>
  <lpage>1109</lpage>
</bibl>

<bibl id="B17">
  <title><p>Sequence context-specific profiles for homology
  searching</p></title>
  <aug>
    <au><snm>Biegert</snm><fnm>A</fnm></au>
    <au><snm>S{\"o}ding</snm><fnm>J</fnm></au>
  </aug>
  <source>Proceedings of the National Academy of Sciences</source>
  <publisher>National Acad Sciences</publisher>
  <pubdate>2009</pubdate>
  <volume>106</volume>
  <issue>10</issue>
  <fpage>3770</fpage>
  <lpage>-3775</lpage>
</bibl>

<bibl id="B18">
  <title><p>Context-dependent optimal substitution matrices</p></title>
  <aug>
    <au><snm>Koshi</snm><fnm>JM</fnm></au>
    <au><snm>Goldstein</snm><fnm>RA</fnm></au>
  </aug>
  <source>Protein Engineering</source>
  <pubdate>1995</pubdate>
  <volume>8</volume>
  <issue>7</issue>
  <fpage>641</fpage>
  <lpage>645</lpage>
</bibl>

<bibl id="B19">
  <title><p>Combining protein evolution and secondary structure.</p></title>
  <aug>
    <au><snm>Thorne</snm><fnm>JL</fnm></au>
    <au><snm>Goldman</snm><fnm>N</fnm></au>
    <au><snm>Jones</snm><fnm>DT</fnm></au>
  </aug>
  <source>Molecular Biology and Evolution</source>
  <publisher>SMBE</publisher>
  <pubdate>1996</pubdate>
  <volume>13</volume>
  <issue>5</issue>
  <fpage>666</fpage>
  <lpage>-673</lpage>
</bibl>

<bibl id="B20">
  <title><p>Markov Models of Amino Acid Substitution to Study Proteins with
  Intrinsically Disordered Regions</p></title>
  <aug>
    <au><snm>Szalkowski</snm><fnm>AM</fnm></au>
    <au><snm>Anisimova</snm><fnm>M</fnm></au>
  </aug>
  <source>PLoS ONE</source>
  <publisher>Public Library of Science</publisher>
  <pubdate>2011</pubdate>
  <volume>6</volume>
  <issue>5</issue>
  <fpage>e20488</fpage>
</bibl>

<bibl id="B21">
  <title><p>Prediction of the stability of protein mutants based on structural
  environment-dependent amino acid substitution and propensity
  tables.</p></title>
  <aug>
    <au><snm>Topham</snm><fnm>CM</fnm></au>
    <au><snm>Srinivasan</snm><fnm>N</fnm></au>
    <au><snm>Blundell</snm><fnm>TL</fnm></au>
  </aug>
  <source>Protein Engineering</source>
  <publisher>Oxford Univ Press</publisher>
  <pubdate>1997</pubdate>
  <volume>10</volume>
  <issue>1</issue>
  <fpage>7</fpage>
  <lpage>-21</lpage>
</bibl>

<bibl id="B22">
  <title><p>A 3D-1D substitution matrix for protein fold recognition that
  includes predicted secondary structure of the sequence</p></title>
  <aug>
    <au><snm>Rice</snm><fnm>DW</fnm></au>
    <au><snm>Eisenberg</snm><fnm>D</fnm></au>
  </aug>
  <source>Journal of Molecular Biology</source>
  <pubdate>1997</pubdate>
  <volume>267</volume>
  <issue>4</issue>
  <fpage>1026</fpage>
  <lpage>1038</lpage>
</bibl>

<bibl id="B23">
  <title><p>FUGUE: sequence-structure homology recognition using
  environment-specific substitution tables and structure-dependent gap
  penalties</p></title>
  <aug>
    <au><snm>Shi</snm><fnm>J</fnm></au>
    <au><snm>Blundell</snm><fnm>TL</fnm></au>
    <au><snm>Mizuguchi</snm><fnm>K</fnm></au>
  </aug>
  <source>Journal of Molecular Biology</source>
  <pubdate>2001</pubdate>
  <volume>310</volume>
  <issue>1</issue>
  <fpage>243</fpage>
  <lpage>257</lpage>
</bibl>

<bibl id="B24">
  <title><p>Modeling $\alpha$-helical transmembrane domains: The calculation
  and use of substitution tables for lipid-facing residues</p></title>
  <aug>
    <au><snm>Blundell</snm><fnm>TL</fnm></au>
    <au><snm>Donnelly</snm><fnm>D</fnm></au>
    <au><snm>Overington</snm><fnm>JP</fnm></au>
    <au><snm>Ruffle</snm><fnm>SV</fnm></au>
    <au><snm>Nugent</snm><fnm>JH</fnm></au>
  </aug>
  <source>Protein Science</source>
  <publisher>Wiley Online Library</publisher>
  <pubdate>1993</pubdate>
  <volume>2</volume>
  <issue>1</issue>
  <fpage>55</fpage>
  <lpage>-70</lpage>
</bibl>

<bibl id="B25">
  <title><p>Environment specific substitution tables improve membrane protein
  alignment</p></title>
  <aug>
    <au><snm>Hill</snm><fnm>JR</fnm></au>
    <au><snm>Kelm</snm><fnm>S</fnm></au>
    <au><snm>Shi</snm><fnm>J</fnm></au>
    <au><snm>Deane</snm><fnm>CM</fnm></au>
  </aug>
  <source>Bioinformatics</source>
  <pubdate>2011</pubdate>
  <volume>27</volume>
  <issue>13</issue>
  <fpage>i15</fpage>
  <lpage>i23</lpage>
</bibl>

<bibl id="B26">
  <title><p>BLOSUM62 miscalculations improve search performance</p></title>
  <aug>
    <au><snm>Styczynski</snm><fnm>MP</fnm></au>
    <au><snm>Jensen</snm><fnm>KL</fnm></au>
    <au><snm>Rigoutsos</snm><fnm>I</fnm></au>
    <au><snm>Stephanopoulos</snm><fnm>G</fnm></au>
  </aug>
  <source>Nature biotechnology</source>
  <publisher>Nature Publishing Group</publisher>
  <pubdate>2008</pubdate>
  <volume>26</volume>
  <issue>3</issue>
  <fpage>274</fpage>
  <lpage>-275</lpage>
</bibl>

<bibl id="B27">
  <title><p>The construction of amino acid substitution matrices for the
  comparison of proteins with non-standard compositions</p></title>
  <aug>
    <au><snm>Yu</snm><fnm>YK</fnm></au>
    <au><snm>Altschul</snm><fnm>SF</fnm></au>
  </aug>
  <source>Bioinformatics</source>
  <publisher>Oxford Univ Press</publisher>
  <pubdate>2005</pubdate>
  <volume>21</volume>
  <issue>7</issue>
  <fpage>902</fpage>
  <lpage>-911</lpage>
</bibl>

<bibl id="B28">
  <title><p>Heads or Tails: A simple reliability check for multiple sequence
  alignments</p></title>
  <aug>
    <au><snm>Landan</snm><fnm>G</fnm></au>
    <au><snm>Graur</snm><fnm>D</fnm></au>
  </aug>
  <source>Molecular Biology and Evolution</source>
  <pubdate>2007</pubdate>
  <volume>24</volume>
  <issue>6</issue>
  <fpage>1380</fpage>
  <lpage>1383</lpage>
</bibl>

<bibl id="B29">
  <title><p>Noisy: Identification of problematic columns in multiple sequence
  alignments</p></title>
  <aug>
    <au><snm>Dress</snm><fnm>A</fnm></au>
    <au><snm>Flamm</snm><fnm>C</fnm></au>
    <au><snm>Fritzsch</snm><fnm>G</fnm></au>
    <au><snm>Grunewald</snm><fnm>S</fnm></au>
    <au><snm>Kruspe</snm><fnm>M</fnm></au>
    <au><snm>Prohaska</snm><fnm>S</fnm></au>
    <au><snm>Stadler</snm><fnm>P</fnm></au>
  </aug>
  <source>Algorithms for Molecular Biology</source>
  <pubdate>2008</pubdate>
  <volume>3</volume>
  <issue>1</issue>
  <fpage>7</fpage>
</bibl>

<bibl id="B30">
  <title><p>An alignment confidence score capturing robustness to guide tree
  uncertainty</p></title>
  <aug>
    <au><snm>Penn</snm><fnm>O</fnm></au>
    <au><snm>Privman</snm><fnm>E</fnm></au>
    <au><snm>Landan</snm><fnm>G</fnm></au>
    <au><snm>Graur</snm><fnm>D</fnm></au>
    <au><snm>Pupko</snm><fnm>T</fnm></au>
  </aug>
  <source>Molecular Biology and Evolution</source>
  <pubdate>2010</pubdate>
  <volume>27</volume>
  <issue>8</issue>
  <fpage>1759</fpage>
  <lpage>1767</lpage>
</bibl>

<bibl id="B31">
  <title><p>Bayesian adaptive sequence alignment algorithms.</p></title>
  <aug>
    <au><snm>Zhu</snm><fnm>J</fnm></au>
    <au><snm>Liu</snm><fnm>J S</fnm></au>
    <au><snm>Lawrence</snm><fnm>C E</fnm></au>
  </aug>
  <source>Bioinformatics</source>
  <pubdate>1998</pubdate>
  <volume>14</volume>
  <issue>1</issue>
  <fpage>25</fpage>
  <lpage>39</lpage>
</bibl>

<bibl id="B32">
  <title><p>{StatAlign: an extendable software package for joint Bayesian
  estimation of alignments and evolutionary trees}</p></title>
  <aug>
    <au><snm>Nov\'ak</snm><fnm>\'A.</fnm></au>
    <au><snm>Mikl\'os</snm><fnm>I.</fnm></au>
    <au><snm>Lyngs\o{}</snm><fnm>R.</fnm></au>
    <au><snm>Hein</snm><fnm>J.</fnm></au>
  </aug>
  <source>Bioinformatics</source>
  <pubdate>2008</pubdate>
  <volume>24</volume>
  <issue>20</issue>
  <fpage>2403</fpage>
  <lpage>2404</lpage>
</bibl>

<bibl id="B33">
  <title><p>Blocks+: a non-redundant database of protein alignment blocks
  derived from multiple compilations.</p></title>
  <aug>
    <au><snm>Henikoff</snm><fnm>S</fnm></au>
    <au><snm>Henikoff</snm><fnm>JG</fnm></au>
    <au><snm>Pietrokovski</snm><fnm>S</fnm></au>
  </aug>
  <source>Bioinformatics</source>
  <publisher>Oxford Univ Press</publisher>
  <pubdate>1999</pubdate>
  <volume>15</volume>
  <issue>6</issue>
  <fpage>471</fpage>
  <lpage>-479</lpage>
</bibl>

<bibl id="B34">
  <title><p>Methods for assessing the statistical significance of molecular
  sequence features by using general scoring schemes</p></title>
  <aug>
    <au><snm>Karlin</snm><fnm>S</fnm></au>
    <au><snm>Altschul</snm><fnm>SF</fnm></au>
  </aug>
  <source>Proceedings of the National Academy of Sciences</source>
  <publisher>National Acad Sciences</publisher>
  <pubdate>1990</pubdate>
  <volume>87</volume>
  <issue>6</issue>
  <fpage>2264</fpage>
  <lpage>-2268</lpage>
</bibl>

<bibl id="B35">
  <title><p>Statistical significance in biological sequence
  analysis</p></title>
  <aug>
    <au><snm>Mitrophanov</snm><fnm>AY</fnm></au>
    <au><snm>Borodovsky</snm><fnm>M</fnm></au>
  </aug>
  <source>Briefings in Bioinformatics</source>
  <pubdate>2006</pubdate>
  <volume>7</volume>
  <issue>1</issue>
  <fpage>2</fpage>
  <lpage>24</lpage>
</bibl>

<bibl id="B36">
  <title><p>Towards a reliable objective function for multiple sequence
  alignments</p></title>
  <aug>
    <au><snm>Thompson</snm><fnm>JD</fnm></au>
    <au><snm>Plewniak</snm><fnm>F</fnm></au>
    <au><snm>Ripp</snm><fnm>R</fnm></au>
    <au><snm>Thierry</snm><fnm>JC</fnm></au>
    <au><snm>Poch</snm><fnm>O</fnm></au>
  </aug>
  <source>Journal of Molecular Biology</source>
  <pubdate>2001</pubdate>
  <volume>314</volume>
  <issue>4</issue>
  <fpage>937</fpage>
  <lpage>951</lpage>
</bibl>

<bibl id="B37">
  <title><p>Automatic assessment of alignment quality</p></title>
  <aug>
    <au><snm>Lassmann</snm><fnm>T</fnm></au>
    <au><snm>Sonnhammer</snm><fnm>ELL</fnm></au>
  </aug>
  <source>Nucleic Acids Research</source>
  <pubdate>2005</pubdate>
  <volume>33</volume>
  <issue>22</issue>
  <fpage>7120</fpage>
  <lpage>7128</lpage>
</bibl>

<bibl id="B38">
  <title><p>Model-based prediction of sequence alignment quality</p></title>
  <aug>
    <au><snm>Ahola</snm><fnm>V</fnm></au>
    <au><snm>Aittokallio</snm><fnm>T</fnm></au>
    <au><snm>Vihinen</snm><fnm>M</fnm></au>
    <au><snm>Uusipaikka</snm><fnm>E</fnm></au>
  </aug>
  <source>Bioinformatics</source>
  <pubdate>2008</pubdate>
  <volume>24</volume>
  <issue>19</issue>
  <fpage>2165</fpage>
  <lpage>2171</lpage>
</bibl>

<bibl id="B39">
  <title><p>Estimating the accuracy of multiple alignments and its use in
  parameter advising</p></title>
  <aug>
    <au><snm>DeBlasio</snm><fnm>D</fnm></au>
    <au><snm>Wheeler</snm><fnm>T</fnm></au>
    <au><snm>Kececioglu</snm><fnm>J</fnm></au>
  </aug>
  <source>Research in Computational Molecular Biology</source>
  <publisher>Springer Berlin Heidelberg</publisher>
  <editor>Chor, Benny</editor>
  <series><title><p>Lecture Notes in Computer Science</p></title></series>
  <pubdate>2012</pubdate>
  <volume>7262</volume>
  <fpage>45</fpage>
  <lpage>59</lpage>
</bibl>

<bibl id="B40">
  <title><p>Alignment metric accuracy</p></title>
  <aug>
    <au><snm>Schwartz</snm><fnm>AS</fnm></au>
    <au><snm>Myers</snm><fnm>EW</fnm></au>
    <au><snm>Pachter</snm><fnm>L</fnm></au>
  </aug>
  <source>arXiv:q-bio/0510052</source>
  <pubdate>2005</pubdate>
</bibl>

<bibl id="B41">
  <title><p>Weak convergence and optimal scaling of random walk {Metropolis}
  algorithms</p></title>
  <aug>
    <au><snm>Roberts</snm><fnm>G. O.</fnm></au>
    <au><snm>Gelman</snm><fnm>A.</fnm></au>
    <au><snm>Gilks</snm><fnm>W. R.</fnm></au>
  </aug>
  <source>Annals of Applied Probability</source>
  <pubdate>1997</pubdate>
  <volume>7</volume>
  <fpage>110</fpage>
  <lpage>-120</lpage>
</bibl>

<bibl id="B42">
  <title><p>On the complexity of multiple sequence alignment</p></title>
  <aug>
    <au><snm>Wang</snm><fnm>L.</fnm></au>
    <au><snm>Jiang</snm><fnm>T.</fnm></au>
  </aug>
  <source>Journal of Computational Biology</source>
  <pubdate>1994</pubdate>
  <volume>1</volume>
  <issue>4</issue>
  <fpage>337</fpage>
  <lpage>348</lpage>
</bibl>

<bibl id="B43">
  <title><p>OXBench: A benchmark for evaluation of protein multiple sequence
  alignment accuracy</p></title>
  <aug>
    <au><snm>Raghava</snm><fnm>GPS</fnm></au>
    <au><snm>Searle</snm><fnm>S</fnm></au>
    <au><snm>Audley</snm><fnm>P</fnm></au>
    <au><snm>Barber</snm><fnm>J</fnm></au>
    <au><snm>Barton</snm><fnm>G</fnm></au>
  </aug>
  <source>BMC Bioinformatics</source>
  <pubdate>2003</pubdate>
  <volume>4</volume>
  <issue>1</issue>
  <fpage>47</fpage>
</bibl>

<bibl id="B44">
  <title><p>Efficient representation of uncertainty in multiple sequence
  alignments using directed acyclic graphs</p></title>
  <aug>
    <au><snm>Herman</snm><fnm>JL</fnm></au>
    <au><snm>Nov\'ak</snm><fnm>A</fnm></au>
    <au><snm>Lyngs\o{}</snm><fnm>R</fnm></au>
    <au><snm>Szab\'o</snm><fnm>A</fnm></au>
    <au><snm>Mikl\'os</snm><fnm>I</fnm></au>
    <au><snm>Hein</snm><fnm>J</fnm></au>
  </aug>
  <source>(submitted)</source>
  <pubdate>2014</pubdate>
</bibl>

</refgrp>
} 

\end{document}